\documentclass[
aps,
prx,
reprint,
superscriptaddress,
nofootinbib,
noeprint,
showkeys,
longbibliography,
amsmath,amssymb,
]{revtex4-2}

\usepackage[UKenglish]{isodate}
\usepackage[ngerman,english]{babel}
\usepackage[T1]{fontenc}
\usepackage{microtype}
\usepackage{hyphenat}
\usepackage{physics}
\usepackage{bm}
\usepackage{bbold}
\usepackage{slashed}

\newcommand{\bigO}{\mathcal{O}}
\newcommand{\sgnb}{\mathrm{sgn}_\mathrm{b}}

\usepackage{comment}
\usepackage[normalem]{ulem} 
\usepackage[dvipsnames]{xcolor}
\usepackage[dvipsnames, svgnames, table]{xcolor}
\definecolor{darkorange}{RGB}{227, 103, 0}
\definecolor{lightblue}{RGB}{93,197, 192}
\definecolor{darkblue}{RGB}{58,129, 158}

\newcommand{\blue}[1]{\textcolor{lightblue}{#1}}

\usepackage{graphicx}
\graphicspath{{graphics/}}
\usepackage[caption=false]{subfig}
\usepackage{adjustbox}
\usepackage{tikz}
\usepackage{quantikz}
\usepackage{wrapfig}
\usepackage[
	raiselinks=true,                 
	bookmarks=true,                     
	bookmarksopenlevel=1,               
	bookmarksopen=true,
	bookmarksnumbered=true,
	hyperindex=true,
	plainpages=false,
	pdfpagelabels=true,
	pdfborder={0 0 0.5},
	colorlinks=true,
	linkcolor={darkblue},
	citecolor={darkblue},
	urlcolor={darkblue},
]{hyperref}
\usepackage[capitalise]{cleveref}
\crefformat{footnote}{#2\footnotemark[#1]#3}

\usepackage[acronym]{glossaries}
\makeglossary
\newacronym{qpe}{QPE}{quantum phase estimation}
\newacronym{qft}{QFT}{quantum Fourier transform}
\newacronym{vqe}{VQE}{variational quantum eigensolver}
\newacronym{nisq}{NISQ}{noisy intermediate scaled quantum}
\newacronym{aa}{AA}{amplitude amplification}
\newacronym{msb}{MSB}{most significant bit}
\newacronym{ces}{CES}{complete eigenpair solver}
\newacronym{pes}{PES}{partial eigenpair solver}
\glsdisablehyper

\begin{document}

\title{Quantum eigenpair solver with minimal sampling overhead}

\author{Sven Danz}
\email{s.danz@fz-juelich.de}
\affiliation{Institute for Quantum Computing Analytics (PGI-12), Forschungszentrum J\"ulich, 52425 J\"ulich, Germany}
\affiliation{Theoretical Physics, Saarland University, 66123 Saarbrücken, Germany}

\begin{abstract}
    The advantage that many quantum algorithms have over their classical counterparts may be lost when the results are extracted as classical data (output problem).
	One example are eigenpair solvers, which encode the eigenpairs in a quantum state.
	Extracting these states results in significant sampling overheads.
	We propose an amplitude-amplification-based post-filtering process that reduces the number of eigenpairs encoded in the final state to a feasible amount.
    Often for practical applications, computing a subset of all eigenpairs is sufficient, which drastically reduces the sampling overhead.
    We show, that our adapted eigenpair solver does not only compete with classical alternatives but outperforms them in terms of memory requirements, runtime, and versatility.
	This makes it an efficient end-to-end quantum algorithm with real-world application in science and engineering.
\end{abstract}

\maketitle

Numerical methods on classical computers enable us to solve numerous problems with greater efficiency than was previously possible.
The majority of these problems can be mapped to mathematical equations that possess a clear solution.
One of the aforementioned equation families is that of eigenpair problems, which find application in a variety of scientific and engineering problems, including the simulation of chemistry, structural, or electrical problems.

The present manuscript is concerned with eigenpair problems of the following form:
\begin{equation}
	\bm{A} \bm{v}_j
	=
	\lambda_j \bm{v}_j.
\end{equation}
Here, $\bm{A}\in\mathbb{R}^{N\times N}$ is a Hermitian matrix with $\bm{v}_j=(v_{1j},\dots,v_{Nj})^T$ and $\lambda_j$ its $j$-th normalized eigenvector and eigenvalue, respectively.
The eigenpair is here defined as $(\lambda_j,\abs{v_{uj}})$: the eigenvalue and a chosen element of the eigenvector\footnote{
	This is a reduced version of the usual definition that contains the full eigenvector.
}.
A complete solution of this problem means finding all eigenpairs of $\bm{A}$ for given $u \in [N]$.
Classically this requires a runtime of $\bigO(N^2s)$ with the QR algorithm for an $s$-sparse matrix, with at most $s$ non-zero elements in each row~\cite{press1992}.
For large matrices, this can result in extensive runtimes, which can either increase expenses or render the problem intractable within acceptable timeframes on state-of-the-art supercomputers.

Solving eigenvalue problems with quantum computers is not a novel concept, and the existing literature can be broadly divided into three main approaches.
The first approach relies on variational quantum algorithms, most notably \gls{vqe}~\cite{peruzzo2014}, which estimates the minimum eigenvalue of a Hamiltonian.
This approach has been well studied for applications in quantum chemistry~\cite{kandalaHardwareefficientVariationalQuantum2017,liVariationalQuantumSimulation2019} and structural dynamics~\cite{sato2023,liu2024}.
\gls{vqe}-based algorithms require only short quantum circuits making them \gls{nisq}-friendly.
However, their efficiency and scalability in terms of runtime are primarily assessed through benchmark studies, making extrapolation to larger and more relevant systems difficult.

The second, but historically earlier, class of eigenvalue solver is based on \gls{qpe}~\cite{kitaevQuantumMeasurementsAbelian1995a,NielsenChuang}.
This method exploits the relationship between the phase of a quantum system’s time evolution operator and the eigenvalues of its Hamiltonian.
\gls{qpe}-based algorithms have also been studied for generalized eigenvalue problems~\cite{parker2020} and for use in modal analysis of elastic structures~\cite{lee2023}.
Unlike \gls{vqe}s, the \gls{qpe}-based approach allows for exact runtime calculations.
However, the presented algorithms are usually not end-to-end, which makes their complexity analysis incomplete.
Furthermore, the aforementioned \gls{qpe} routines solve only pure eigenvalue problems and require modifications for the extraction of the corresponding eigenvector.

A more recent eigenvalue solver combines the Hadamard test with classical signal processing methods~\cite{lin2022, dutkiewicz2022, wang2023}.
Besides the Hadamard test, it only requires access to quantum time evolution, yielding shorter circuits than the \gls{qft}-based \gls{qpe}, which makes it \gls{nisq}-friendlier.
Its total runtime is similar to that of textbook \gls{qpe}~\cite{NielsenChuang}.
Recent updates address previous limitations regarding the spectral gap requirements~\cite{ding2024}.

In this manuscript, we demonstrate how to solve the full eigenpair problem using quantum computing to achieve exponential memory savings and end-to-end runtime improvements over classical methods.
We first review a recently introduced \gls{qpe}-based complete quantum eigenpair solver, which provides exponential memory advantages but often falls short in runtime compared to classical algorithms~\cite{danz2024response}. 
Building on this, we add amplitude amplification to focus on a relevant subset of the problem, leading to a quantum routine that closes this gap and ultimately outperforms classical alternatives in both memory and runtime.

\section{Complete eigenpair solver}
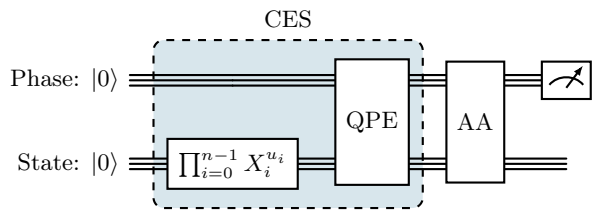
\begin{figure}
	\centering
	\begin{adjustbox}{max width=\linewidth}
		\begin{quantikz}[wire types={b,b},classical gap=0.07cm]
			\lstick{Phase: $|0\rangle$} &\gategroup[
				2,
				steps=2,
				style={
						dashed,
						rounded corners,
						fill=darkblue!20,
						inner xsep=2pt
					},
				background,
				label style={
						label position=above,
						anchor=south,
						yshift=-0.15 cm
					}
			]{CES}&\gate[2]{\mathrm{QPE}}&\gate[2]{\mathrm{AA}}&\meter{}
			\\
			\lstick{State: $|0\rangle$} &\gate{\prod_{i=0}^{n-1}X^{u_i}_i}&&&
		\end{quantikz}
	\end{adjustbox}
	\caption{
		\Acrfull{pes} based on \acrfull{qpe}~\cite{danz2024response}.
		The original \acrfull{ces} consists only of the encoding of a computational basis state via NOT gates followed by the \gls{qpe}.
		This computes all eigenpairs in parallel.
		We extend the routine by an \acrfull{aa} reducing the number of solutions.
	}
	\label{fig:eigenpair_solver}
\end{figure}

The algorithm proposed in Ref.~\cite{danz2024response} consists of three steps: The amplitude encoding of the eigenvector in $n=\lceil\log_2 N\rceil$ qubits, an $\bm{A}$ dependent \gls{qpe} preparing the eigenvalues $\lambda_j$ in a secondary register, and a measurement of the secondary register at the end (see \cref{fig:eigenpair_solver} minus the \acrshort{aa} gate).

The direct implementation of an amplitude encoding $\ket{v_j}$ of an eigenvector is impossible if $v_j$ is unknown, which is why they propose the preparation of a computational state $\ket{u}$ instead
\begin{equation}
	\ket{u}
	=
	\sum_{j=1}^N v_{uj}\ket{v_j}.
	\label{eq:init_state}
\end{equation}
Here, $\{\ket{v_1},\dots,\ket{v_N}\}$ forms a complete eigenbasis.
The encoding of $\ket{u}$ prepares all eigenstates without knowing any of them, while requiring at most one iteration of NOT gates denoted by $X$.

It follows the \gls{qpe} for the unitary $V$ with the properties
\begin{equation}
	V\ket{\mu_{\pm}^{(j)}}
	=
	e^{\pm i f(\lambda_j)}\ket{\mu_{\pm}^{(j)}},
\end{equation}
here, $f$ is a function of which we know the inverse $f^{-1}$ and $|\mu_{\pm}^{(j)}\rangle$ is a new basis
\begin{equation}
	\ket{v_j}
	=
	\frac{1}{\sqrt{2}}\left(
	\ket{\mu_{+}^{(j)}}
	+\ket{\mu_{-}^{(j)}}
	\right).
\end{equation}
Applying \gls{qpe} to \cref{eq:init_state}, extended by an $m$-qubit register, yields
\begin{multline}
	\mathrm{QPE}\sum_{j=1}^N v_{uj}\ket{0}^{\otimes m}\ket{v_j}
	=
	\\
	\frac{1}{\sqrt{2}}\sum_{j=1}^N v_{uj} \left(
	\ket{\varphi_j}\ket{\mu_{+}^{(j)}}
	+\ket{-\varphi_j}\ket{\mu_{-}^{(j)}}
	\right).
	\label{eq:qpe}
\end{multline}
Here, we simplified the final state by removing amplitudes close to zero (see Ref.~\cite{danz2024response} for more details).
The eigenvalues are hidden in $\varphi_j = 2^mf(\lambda_j)/2\pi$ binarily encoded in the $m$ ancilla qubits.
The measurement of them lets the superposition collapse and returns $\varphi_j$ with probability $\abs{v_{uj}}^2/2$.
Knowing $f^{-1}$ allows extracting $\lambda_j$.

Extracting all eigenvalues paired with their probabilities, requires a large sample number.
A total of $\bigO(\delta^{-2}\ln(N\zeta^{-1}))$ samples are required to guarantee an estimation of $\abs{v_{uj}}^2$ within tolerance $\delta$ and failure probability of $\zeta$ for $N$ eigenvalues.
However, the eigenvector $\bm{v}_j$ is normalized which means that the average $\abs{v_{uj}}^2$ and with it the required tolerance $\delta = \gamma \abs{v_{uj}}^2=\bigO( \gamma N^{-1})$ decreases with $N$.
$\gamma$ denotes the relative tolerance without hidden $N$ dependency.
This gives a sampling overhead of
\begin{equation}
	t_\mathrm{s}
	=
	\bigO\left(
	\frac{N^2}{\gamma^2}\ln (\frac{N}{\zeta})
	\right),
\end{equation}
next to the runtime of the \gls{qpe}.
The described method is only competitive with classical algorithms if one has knowledge that allows us to choose a $u$ with large $\abs{v_{uj}}=\bigO(1)$.

This algorithm computes the eigenpairs for a complete superposition of all eigenstates $\ket{v_j}$ of our matrix $\bm{A}$.
Hence, we refer to it as \gls{ces}.

\section{Partial eigenpair solver}

For many problems it is sufficient to focus only on a small subset of all eigenpairs.
We call the corresponding routine \acrfull{pes}.
To achieve this, we suggest the addition of an \acrfull{aa} to the original eigenpair solver (cf. \cref{fig:eigenpair_solver} and Refs.~\cite{grover1996,brassard2002}).
The \gls{aa} can be tuned to return mostly eigenstates with eigenvalues within a given interval $[\lambda_\mathrm{l},\lambda_\mathrm{r})$
\begin{multline}
	\mathrm{AA}\sum_{j=1}^N v_{uj}\ket{v_j}
	\\
	=
	c\sum_{j\in\mathcal{J}}  v_{uj}\ket{v_j}
	+\sqrt{1-c^2}\sum_{j\notin\mathcal{J}} v_{uj}\ket{v_j},
\end{multline}
where $\mathcal{J}=\{j:\lambda_j\in[\lambda_\mathrm{l},\lambda_\mathrm{r})\}$ and $c$ is an \gls{aa} dependent coefficient.
This increases the probability of measuring a state $\ket{v_j}:j\in\mathcal{J}$ to $\abs{cv_{uj}}^2$.
Here, $c$ should be chosen such that $\abs{cv_{uj}}^2$ and its tolerance $\delta=\gamma \abs{cv_{uj}}^2$ are independent of $N$.
This is given for $\abs{c}^2 = \bigO(Nk^{-1})$, where $k\ll N$ is the number of eigenvalues within $[\lambda_\mathrm{l},\lambda_\mathrm{r})$.
In this case, the sampling overhead reads
\begin{equation}
	t_\mathrm{s}'
	=
	\bigO\left(
	\frac{k^2}{\gamma^2}\ln (\frac{k}{\zeta})
	\right).
\end{equation}

Small enough intervals $[\lambda_\mathrm{l},\lambda_\mathrm{r})$ (and with it $k$) make this negligible in contrast to the runtime of the \gls{pes}
\begin{equation}
	t_\mathrm{PES}
	=
	t_\mathrm{CES}
	+t_\mathrm{AA}.
\end{equation}
$t_\mathrm{CES}$ and $t_\mathrm{AA}$ denote the runtimes of the \gls{ces} and the \gls{aa} respectively.

\section{Eigenvalue based amplitude amplification}
The need for the amplification of single states in a quantum superposition is far from novel and well studied in the field of \gls{aa}~\cite{brassard2002}.
Its oldest example is Grover's search algorithm~\cite{grover1996} with quadratic speed up.
Generalized in the context of quantum signal processing, \gls{aa} can be used to amplify the amplitudes of a given quantum superposition to match versatile polynomials~\cite{martyn2021, lin2020}.

The implementation of \gls{aa} requires two reflection operators \cite{grover1996,brassard2002}.
The first $R_\mathrm{G}$ reflects around the set of \emph{good} states $\{\ket{v_j}: j\in\mathcal{J}\}$ and the second $R_\mathrm{state}$ reflects around the original state.
We identify the \emph{good} states by their corresponding eigenvalues $\lambda_j\in[\lambda_\mathrm{l}, \lambda_\mathrm{r})$ to which we get access via the phase register of the \gls{qpe} (see \cref{eq:qpe}).

\begin{figure}
	\centering
	\begin{adjustbox}{max width=\linewidth}
		\begin{quantikz}[wire types={q,b,q},classical gap=0.07cm]
			\lstick{\blue{$|0\rangle$}} &\gategroup[
				3,
				steps=3,
				style={
						dashed,
						rounded corners,
						fill=darkblue!30,
						inner xsep=2pt
					},
				background,
				label style={
						label position=above,
						anchor=south,
						yshift=-0.15 cm
					}
			]{$U_\mathrm{SAS}$}&&\gate[2]{\mathrm{qcSUB}_{\varphi_\mathrm{r}}}\gateoutput{msb.}&\ctrl{2}&\gate[3]{U_\mathrm{SAS}^\dagger}&\rstick{\blue{$|0\rangle$}}
			\\
			\lstick{ $\ket{\varphi_j}$} &\gate[2]{\mathrm{qcSUB}_{\varphi_\mathrm{l}}}&\gate{\mathrm{qcADD}_{\varphi_\mathrm{l}}}&&&&\rstick{ $\ket{\varphi_j}$}
			\\
			\lstick{\blue{$|0\rangle$}} &\gateoutput{msb.}&\gate{X}&&\control{}&&\rstick{\blue{$|0\rangle$}}
		\end{quantikz}
	\end{adjustbox}
	\caption{
	Reflection $R_G$ around the \emph{good} states $\ket{v_j},\forall j\in\mathcal{J}$.
	We use the quantum adder $\mathrm{qcADD}_l$ and subtractor $\mathrm{qcSUB}_l$ to compare $\varphi_j$ with the left and right limits defining the \emph{good} states $[\varphi_\mathrm{l},\varphi_\mathrm{r})$ before combining the results with a controlled $Z$ gate.
	At last, all qubits are returned into their original bases with the Hermitian conjugate of the first three steps leaving only the global phase.
	}
	\label{fig:test}
\end{figure}
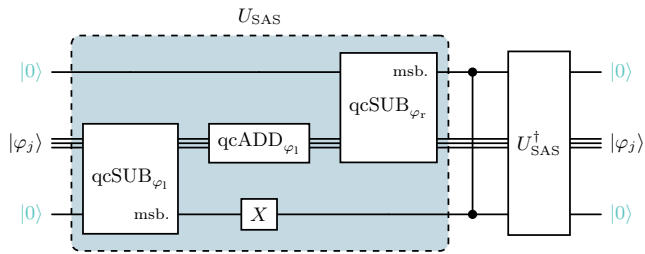

The test $\lambda_j\in[\lambda_\mathrm{l},\lambda_\mathrm{r})$ or as a substitute $\varphi_j\in[\varphi_\mathrm{l},\varphi_\mathrm{r})$ with $\varphi_\mathrm{l/r}=2^m\abs{f(\lambda_\mathrm{l/r})}/2\pi$, can be implemented via a series of quantum-classical\footnote{\emph{Quantum-classical} means that one summand is stored in a quantum register and the other in a classical register.} adders $\mathrm{qcADD}_l$ and their inverse subtractors $\mathrm{qcSUB}_l=\mathrm{qcADD}_l^\dagger$ as shown in \cref{fig:test} (see Refs.~\cite{vedral1996,beckman1996,cuccaro2004,gidney2018,thomsen2008,vanmeter2005,zalka1998,draper2004,Ruiz-Perez2017,draper2000} for more details about quantum adders).
It reads the $m$-qubit phase register $\ket{\varphi_j}$ from \cref{fig:eigenpair_solver} and subtracts $\varphi_\mathrm{l}$ from $\ket{0,\varphi_j}$ where one additional qubits is treated as \gls{msb}.
This gives us
\begin{multline}
	\ket{\varphi_j-\varphi_\mathrm{l}\mod 2^{m+1}}
	\\
	=
	\ket{\sgnb(\varphi_j-\varphi_\mathrm{l}),\varphi_j-\varphi_\mathrm{l}\mod 2^{m}}.
\end{multline}
Here, we introduce the binary sign defined by $\sgnb(x\geq0)=0$, $\sgnb(x<0)=1$.
Now, the first qubit contains the opposite of the desired $\lnot\sgnb(\varphi_j-\varphi_\mathrm{l})$ (i.e.~$\varphi_j\geq\varphi_\mathrm{l}$ for $\varphi_j\in[\varphi_\mathrm{l},\varphi_\mathrm{r})$).
A simple NOT gate $X$ solves this and we have
\begin{equation}
	\ket{\lnot\sgnb(\varphi_j-\varphi_\mathrm{l}),\varphi_j-\varphi_\mathrm{l}\mod 2^{m}}.
\end{equation}
An adder returns the main register into its original state
\begin{equation}
	\ket{\lnot\sgnb(\varphi_j-\varphi_\mathrm{l}),\varphi_j}.
\end{equation}
Next, we use a second ancilla qubit as \gls{msb} of the main register and subtract $\varphi_\mathrm{r}$
\begin{multline}
	\ket{\lnot\sgnb(\varphi_j-\varphi_\mathrm{l})}\ket{\varphi_j-\varphi_\mathrm{r}\mod 2^{m+1}}
	\\
	=
	\ket{\lnot\sgnb(\varphi_j-\varphi_\mathrm{l})}
	\\
	\times\ket{\sgnb(\varphi_j-\varphi_\mathrm{r}),\varphi_j-\varphi_\mathrm{r}\mod 2^{m}}.
\end{multline}
The two binary signs are to combine with a controlled $Z$ gate in a global phase
\begin{multline}
	(-1)^{
		\lnot\sgnb(\varphi_j-\varphi_\mathrm{l})
		\times\sgnb(\varphi_j-\varphi_\mathrm{r})
	}
	\ket{\lnot\sgnb(\varphi_j-\varphi_\mathrm{l})}
	\\
	\times\ket{\sgnb(\varphi_j-\varphi_\mathrm{r}),\varphi_j-\varphi_\mathrm{r}\mod 2^{n}}.
\end{multline}
The global sign is either $-1$ for  $\varphi_j\in[\varphi_\mathrm{l},\varphi_\mathrm{r})$ or $1$ for  $\varphi_j\notin[\varphi_\mathrm{l},\varphi_\mathrm{r})$.
The Hermitian conjugate of the two adders, the one subtractor, and the NOT gate returns the main register and the two ancilla qubits into their original state except for the global sign
\begin{equation}
	(-1)^{
		\lnot\sgnb(\varphi_j-\varphi_\mathrm{l})
		\times \sgnb(\varphi_j-\varphi_\mathrm{r})
	}\ket{0}^{\otimes 2}\ket{\varphi_j}.
\end{equation}
All those steps combined form the reflection $R_\mathrm{G}$.

The second reflection $R_\mathrm{state}$ (around the original state, i.e.~the state after the \gls{ces}, cf. \cref{fig:eigenpair_solver}) requires the \gls{ces} plus a reflection along the zero state.
\begin{equation}
	R_\mathrm{state}
	=
	\mathrm{CES}(2\ketbra{0}{0}^{\otimes m+n}-I_{m+n})\mathrm{CES}^\dagger.
\end{equation}
$I_{n}$ denotes the $n$ dimensional identity matrix.
The alternation of those two routines form the $t$-step \gls{aa}
\begin{equation}
	\mathrm{AA}
	=
	(R_\mathrm{state}R_\mathrm{G})^t.
\end{equation}

In worst case we start with a probability of $\sim k/N$ in the good states.
The amplification requires $t=\bigO(\sqrt{N/k})$ cycles~\cite{grover1996,brassard2002} and with it repetitions of the \gls{ces} leading to a \gls{pes} runtime of
\begin{multline}
	t_\mathrm{PES}
	=
	\bigO\left(
	\sqrt{\frac{N}{k}}t_\mathrm{CES}
	\right)
	\\
	=
	\bigO\left(
	\sqrt{\frac{N}{k}}s\norm{A}_\mathrm{max} \max\left(
		\frac{1}{\varepsilon},
		\frac{k}{\gamma \Delta_\lambda^{(u,\mathcal{J})}}
		\right)
	\right).
	\label{eq:t_pes}
\end{multline}
In the second step we replace the runtime $t_\mathrm{CES}$ with the query complexity derived in Ref.~\cite{danz2024response}, were, $\norm{A}_\mathrm{max}=\max_{uv}\abs{A_{uv}}$, $\varepsilon$ denotes the eigenvalue tolerance and $\Delta^{(u,\mathcal{J})}_\lambda$ the minimum gap between eigenvalues $\lambda_j\in[\lambda_\mathrm{l},\lambda_\mathrm{r})$ for which $v_{uj}\neq 0$.

\section{Comparison with classical algorithms}
We modified the eigenpair solver proposed in Ref.~\cite{danz2024response} to be restricted to a small subset of eigenpairs.
Therefore, instead of comparing our method to the state-of-the-art method for complete eigenpair calculations, the QR algorithm ($t_\mathrm{QR}=\bigO(N^2s)$)~\cite{press1992,Demmel1997,GolubVanLoan2013,TrefethenBau1997}, we focus on classical partial eigenpair solvers that beats the QR algorithm for small eigenpair subsets.

Starting with the power method, a routine based on the repetitive multiplication of the matrix $A$ with a random vector $\tilde{\bm{v}}=\sum_{j=1}^{N}c_j\bm{v}_j$.
The matrix amplifies the contribution of the largest eigenvector $\bm{v}_1$ giving a good approximation of the same after a few iterations.
This method is mainly dominated by the sparse matrix-vector multiplication requiring $\bigO(Ns)$ basic arithmetic operations.
The error of this method $\varepsilon$ decreases with an increasing number of iterations $l$
\begin{equation}
	\varepsilon
	\propto
	\left(
	\frac{\lambda_2}{\lambda_1}
	\right)^l,
\end{equation}
where $\lambda_1>\lambda_2$ are the two largest eigenvalue.
This yields in a total runtime of
\begin{equation}
	t_\mathrm{Power}
	=
	\bigO\left(
	Ns k\frac{
		\log\varepsilon
	}{
		\log(\frac{\lambda_2}{\lambda_1})
	}
	\right).
	\label{eq:t_pow}
\end{equation}
Here, we find an additional factor of $k$ as this method finds only one eigenpair at a time.
In order to find the second-largest eigenvalue $\lambda_2$ one has to damp the largest eigenvalue in the matrix $A\to A-\lambda_1 \bm{v}_1\bm{v}_1^\dagger$.
This should be repeated until one has the $k$ largest eigenvalues.
We recommend Refs.~\cite{Demmel1997,GolubVanLoan2013} for more details about the power method.

\begin{figure*}
	\centering
	\includegraphics[width=15cm]{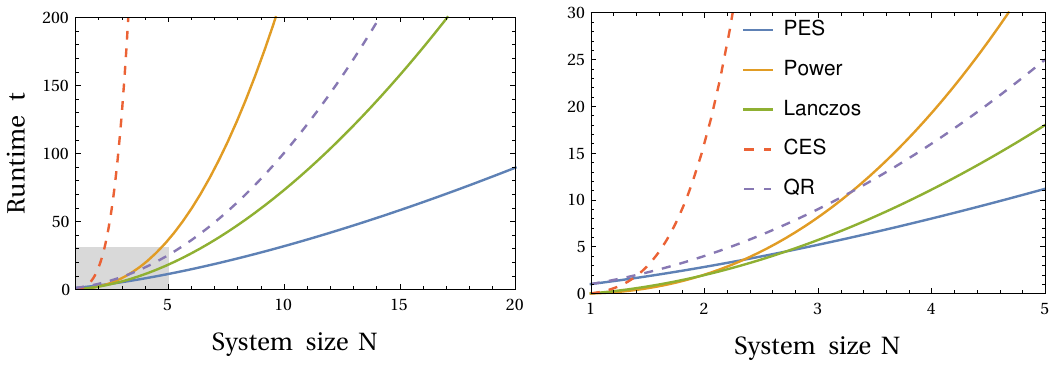}
	\caption{
		Runtime $t$ comparison for the \acrfull{pes}, the power method, the Lanczos method, the \acrfull{ces}, and the QR algorithm under worst conditions (i.e.~$\varepsilon, \Delta_\lambda^{(u,\mathcal{J})},\delta,(1-\lambda_2/\lambda_1)\propto N^{-1}$ with $N$ the matrix size).
		All other parameters are set to 1.
		The plot on the right shows the regime illustrated in gray in the left plot.
	}
	\label{fig:scaling}
\end{figure*}

Compared with \cref{eq:t_pes}, our \gls{pes} outperforms the classical routine quadratically in $N$.
However, here again, we are interested in the worst-case scenario in which the eigenvalues are spread homogeneously and with it $\varepsilon=\bigO(N^{-1})$ and $\lambda_2/\lambda_1=\bigO(1-N^{-1})$.
It is still difficult to compare \cref{eq:t_pow,eq:t_pes}, which is why we show their scaling graphically in \cref{fig:scaling}.
It is clear that our \gls{pes} outperforms the power method for non-trivial problem size.
Furthermore, the non-optimal QR algorithm is also faster than the power method due to its strong scaling with $\lambda_2/\lambda_1$.
We should further mention, that we can not choose an arbitrary interval $[\lambda_\mathrm{l},\lambda_\mathrm{r})$ with the power method.
The power method is limited to the $k$ largest eigenvalues if we have no fast access to $A^{-1}$.

The power method is beaten by multiple other classical routines nowadays.
One of them is the Lanczos algorithm, which is based on the power method.
It is able to find a subset of all eigenpairs fast by forming a Krylov subspace.
This means we replace $A\in \mathbb{R}^{N\times N}$ with a smaller tridiagonal matrix $T\in\mathbb{R}^{l\times l}$ that shares the largest and smallest eigenvalues with $A$ up to an error.
This is also an iterative process with an average error for the largest eigenvalue that is bounded by~\cite{kuczynski1992}
\begin{equation}
	\varepsilon
	=
	\bigO\left(
	\frac{\log^2 N}{l^2}
	\right).
	\label{eq:err_lanczos}
\end{equation}
Here, the error is averaged over all possible initial vectors $\tilde{\bm{v}}$.
Similar to the power method, this requires matrix-vector multiplication ($\bigO(Ns)$) in each iteration, leading to a runtime of
\begin{equation}
	t_\mathrm{Lanczos}
	=
	\bigO\left(
	Ns \frac{
		\log N
	}{
		\sqrt{\varepsilon}
	}
	\right).
	\label{eq:t_lanczos}
\end{equation}
We omit the additional steps necessary to find all eigenpairs of $T$ using e.g. the QR as their runtime $t_\mathrm{QR} =\bigO(l^2)$ scales slower than the space reduction.
The classical Lanczos algorithm is described in detail in Refs.~\cite{Demmel1997,TrefethenBau1997,GolubVanLoan2013}.

The worst case scenario means again, that we have $\varepsilon =\bigO(N^{-1})$, which leads to runtime similar to \cref{eq:t_pes} up to logarithmic terms.
We show a comparison of both in \cref{fig:scaling} with our \gls{pes} surpassing the Lanczos method.
One has to keep in mind that the error bound in \cref{eq:err_lanczos} is only valid for the largest eigenvalue and increases for the second or even $k$-th largest eigenvalue.
This brings us to the big disadvantage of the Lanczos method compared to ours, which is its limitation to the extreme eigenvalues in contrast to our algorithm.

In terms of memory requirements, our method outperforms all mentioned classical routines exponentially.
This is because the classical methods require the storage of sparse matrices, i.e. at least $\bigO(Ns)$ bits.
The method proposed by us, assuming oracle access to $\bm{A}$, stores the matrices in the Hilbert space of a logarithmic number of qubits~\cite{danz2024response} and therefore outperforms their classical alternatives.

\vspace{0mm}

\section{Conclusion}
\label{sec:conclusion}
We propose a \acrfull{pes} based on the \acrfull{ces} introduced in Ref.~\cite{danz2024response}, that computes only a small subset of eigenpairs of a matrix.
This subset is sufficient for a large family of use cases.
The main addition is an amplitude-amplification-based eigenvalue filter, that improves the runtime polynomial.

The proposed \gls{pes} mitigates the dependency of the computational complexity on the desired tolerances and allows us to compete with and outperform classical alternatives in terms of memory requirements, runtime and versatility in a worst-case comparison.
With this, we deliver an end-to-end quantum algorithm with unmatched performance and real-world applications in science and engineering.

Improvements of the algorithm could be obtained by combining it with quantum amplitude estimation instead of amplitude amplification alone.
Finally, we can only provide upper limits for the runtime, postponing the description of its full potential until benchmarking tests on fault-tolerant quantum hardware are possible.

\section*{Acknowledgements}
\label{sec:acknowledgements}
SD were funded by the Bundesministerium für Wirtschaft und Klimaschutz (BMWK, Federal Ministry for Economic Affairs and Climate Action) in the quantum computing enhanced service ecosystem for simulation in manufacturing project (QUASIM, Grand No. 01MQ22001A).

\bibliography{lit.bib}

\end{document}